\documentclass[prd,twocolumn,a4paper,superscriptaddress]{revtex4}
\usepackage{graphicx,amssymb}

\begin{document}
%My commands
\newcommand{\be}{\begin{equation}}
\newcommand{\ee}{\end{equation}}
\newcommand{\bq}{\begin{eqnarray}}
\newcommand{\eq}{\end{eqnarray}}
\newcommand{\bsq}{\begin{subequations}}
\newcommand{\esq}{\end{subequations}}
\newcommand{\bc}{\begin{center}}
\newcommand{\ec}{\end{center}}
\newcommand {\R}{{\mathcal R}}
\newcommand{\al}{\alpha}
\newcommand\lsim{\mathrel{\rlap{\lower4pt\hbox{\hskip1pt$\sim$}} \raise1pt\hbox{$<$}}}
\newcommand\gsim{\mathrel{\rlap{\lower4pt\hbox{\hskip1pt$\sim$}} \raise1pt\hbox{$>$}}}

\title{Dark matter from cosmic defects on galactic scales?}
\author{N. Guerreiro} 
\email[Electronic address: ]{c0770079@alunos.fc.up.pt}
\affiliation{Centro de Astrof\'{\i}sica da Universidade do Porto, Rua das Estrelas, 4150-762 Porto, Portugal}
\affiliation{Departamento de Matem\'atica Aplicada da Faculdade de Ci\^encias da Universidade do Porto, Rua do Campo Alegre 687, 4169-007, Porto, Portugal}
\author{P.P. Avelino} 
\email[Electronic address: ]{ppavelin@fc.up.pt} 
\affiliation{Centro de F\'{\i}sica do Porto, Rua do Campo Alegre 687, 4169-007 Porto, Portugal} 
\affiliation{Departamento de F\'{\i}sica da Faculdade de Ci\^encias da Universidade do Porto, Rua do Campo Alegre 687, 4169-007 Porto, Portugal} 
\author{J.P.M. de Carvalho}
\email[Electronic address: ]{mauricio@astro.up.pt}
\affiliation{Centro de Astrof\'{\i}sica da Universidade do Porto, Rua das Estrelas, 4150-762 Porto, Portugal}
\affiliation{Departamento de Matem\'atica Aplicada da Faculdade de Ci\^encias da Universidade do Porto, Rua do Campo Alegre 687, 4169-007, Porto, Portugal}
\author{C.J.A.P. Martins} 
\email[Electronic address: ]{Carlos.Martins@astro.up.pt}
\affiliation{Centro de Astrof\'{\i}sica da Universidade do Porto, Rua das Estrelas, 4150-762 Porto, Portugal}
\affiliation{DAMTP, University of Cambridge, Wilberforce Road, Cambridge CB3 0WA, United Kingdom} 
\date{26 July 2008}
\begin{abstract}
We discuss the possible dynamical role of extended cosmic defects on galactic scales, specifically focusing on the possibility that they may provide the dark matter suggested by the classical problem of galactic rotation curves. We emphasize that the more standard defects (such as Goto-Nambu strings) are unsuitable for this task, but show that more general models (such as transonic wiggly strings) could in principle have a better chance. In any case, we show that observational data severely restricts any such scenarios.
\end{abstract}
\pacs{98.80.Cq, 11.27.+d}
\keywords{Cosmic strings; Dark matter}
\maketitle

\section{\label{intr}Introduction}

Symmetry breaking phase transitions in the early universe are expected to have produced networks of topological defects \cite{Vilenkin}. The possible roles of these defect networks in key cosmological scenarios will depend both on the type of defect considered and on the corresponding dynamics. For example, if defects are to significantly contribute to the dark energy, they should have a negative equation of state. If so the best possible situation is that of a frustrated network, in which case
\begin{equation}
w\equiv\frac{p}{\rho}=-\frac{N}{3}
\end{equation}
where $N$ is the defect's spatial dimension ($N=1,2$ respectively for cosmic strings and domain walls, while $N=3$ corresponds to a cosmological constant). 

It is clear that only the cases with $N=2$ and $N=3$ can lead to the recent acceleration of the Universe. A crucial property shared by these two cases is that the ratio between the dark energy density and the background density grows rapidly with time, with dark energy being dynamically dominant only around today. The defect network should therefore have a present density close to critical ($\Omega^0_{de}\sim1$), but compatibility with other cosmological observables requires it to have a characteristic scale several orders of magnitude below the horizon $\xi \ll H^{-1}$.

Defects can also act as seeds for structure formation \cite{Silk}. This may be the case if the ratio between the average defect energy density and the background density is approximately a constant and the characteristic scale of the network is roughly proportional to the Hubble radius ($\xi \propto H^{-1}$). Moreover, the defect fluctuations on the Hubble scale must be small ($\delta\lsim 10^{-5}$), and if the characteristic scale of the defects is of the order of the Hubble scale then it follows that their average energy density must also be very small ($\Omega_{def}\lsim 10^{-5}$) for consistency with cosmic microwave background anisotropies.  It is well known that a scaling cosmic string network has all these properties \cite{Vilenkin}. Notice that the required network properties are mutually incompatible: although there are unified dark energy scenarios in which dark matter and dark energy are described by a single entity, it is not possible for a given defect network to simultaneously be the dark energy and act as a seed for structure formation. 

The possible role of domain walls as dark energy candidates has been investigated in detail in \cite{IDEAL1,IDEAL3} where it was shown that the dynamics of realistic domain wall networks appears to be incompatible with a dark energy role. Here we take a closer look at the role of defects on smaller scales. Specifically we will be mostly interested in kiloparsec (that is, galactic) scales. In particular, in doing this we will also explore another possibility that has been recently put forward \cite{Redington,Alex}, \textit{viz.} that cosmic strings could be the dark matter and explain the observed flat rotation curves of spiral galaxies.

\section{\label{Dark}Dark matter}

The oldest evidence for the existence of dark matter comes from the pioneering work of Zwicky \cite{Zwicky} in the 1930s, on the velocity dispersion of galaxies in clusters. On the other hand the evidence for the presence of dark matter on galactic scales comes from the behavior of circular velocities of stars and gas as a function of their distance from the galactic center \cite{Rubin,Faber,Thonnard}, and has been known since the 1970s.

Measurements are normally carried out by combining observations of the 21 cm hydrogen line with optical surface photometry. The observed rotation curves usually exhibit a characteristic flat behavior at large distances, around and well beyond the edge of the visible disks. If one assumes the validity of Newtonian gravity, the circular velocity is given by
\begin{equation}
v(r)^2=\frac{GM(r)}{r}\,.
\end{equation}
In this framework, the observation of a constant large-scale velocity is usually explained by the existence of a dark mater halo, with a mass $M \propto r$. There is also evidence for dark matter in elliptical galaxies, in particular coming from strong gravitational lensing measurements \cite{Koopmans}.

However, the dark matter scenario presents some puzzles. There is no particle, predicted by the standard model of particle physics, whether elementary or composite, that can account for the amount of dark matter required by modern cosmological and astrophysical observations. Big bang nucleosynthesis determines the total amount of baryonic matter present in the Universe to be $ 0.017\leq\Omega_{B} h^{2}\leq 0.024 $ \cite{Brian} , where $ h\equiv H_{0}/100 Kms^{-1}Mpc^{-1}\sim 0.7 $, while the 5-year WMAP data \cite{komatsu,dunkley} implies $ \Omega_{B} h^{2}= 0.02273 \pm 0.00062 $, with the density of luminous matter being $ \Omega_{lum}\simeq 0.0024h^{-1} $\cite{Fukugita}. We may therefore conclude that some of the baryons are dark, although the WMAP data and nucleosynthesis constraints clearly show that they are not sufficient to account for most of the dark matter needed. At the same time, these results rule out the so-called dark bodies (such as black holes, planets or brown dwarfs) as primary dark matter candidates. Therefore, the conclusion that one may draw is that most of dark matter has a nonbaryonic nature.

Given the success of the particle physics standard model, this could therefore point to the existence of new physics beyond it. Indeed, this is one of the key motivations for the forthcoming LHC experiments at CERN. Whatever this new physics may be it should come with new particles---one or more of which could be responsible for the dark matter---in fact, there is no shortage of candidates \cite{Bertone}.

Alternatively, the new physics could be in the gravitational sector. For example, the modified Newtonian dynamics (MOND) \cite{Milgrom} is an empirically-based modification of Newtonian gravity in the limit of low accelerations which describes reasonably well the dynamics in most galaxies and other larger astrophysical systems. A tensor-vector-scalar gravity (TeVeS) theory has also recently been proposed by Bekenstein \cite{Bekenstein:2004ne} which reduces to MOND in the appropriate limit, but it can also be used to make cosmological predictions or to describe gravitational lensing, which could not be addressed by MOND. Although these and other theories \cite{Moffat11} present some successful results on different scales, they face several challenges, such as those posed by the latest cosmic microwave background results \cite{Skordis:2005xk} or the ones implied by the inferred dynamics of the bullet cluster \cite{Clowe}.

In what follows we will explore a new mechanism where the flatness of observed rotation curves in spiral galaxies is due to the presence of cosmic strings. Doing so in fact leads us to a broader analysis of the possible roles of defects on galactic scales.

\section{\label{Strings}Cosmic strings and spiral galaxies}

In order to explain the galaxy rotation curves one would need a cosmic string to go through the center of each galaxy. The string network would need to have a characteristic scale of the order of the average intergalaxy distance $\sim 1 \rm Mpc$ and the string velocities  would have to be very small (basically equal to the velocity of the galaxy). Such scenarios can in principle exist, although they will in general lead to string-dominated universes.

The first such example, using simple Goto-Nambu strings, was proposed by Kibble \cite{Kibble}. Such a string-dominated universe would have $\xi\propto a^{3/2}\propto t$ (neglecting energy losses to loops), and the present-day ratio $(\xi/t)_0=\sqrt{30G\mu}$ then corresponds to $\xi\sim1-30\, {\rm kpc}$. Kibble himself points out that such a scenario relies on a particular set of (somewhat unphysical) phenomenological parameters. Notice that this is no solution to the dark matter problem: long strings could hardly be bound in galaxies---indeed, the analysis also neglects the dynamical effect of the string velocities. 

Much more recently, Alexander \cite{Alex} suggests that dark matter stems from a frozen (rigid) string network, which would provide an accretion mechanism at recombination and thread galaxies, leading to flat rotation curves. Note that a viable mechanism requires that strings explain the dark matter on both large (cosmological) and small (galactic/cluster) scales. Clearly this scenario requires a rigid network ($v \sim 0$) whose strings have wiggles (for an attractive radial force leading to accretion), and obviously dark matter should have an equation of state $w\sim0$. These are mutually incompatible for Goto-Nambu strings, since $v^2 \sim 1/2$ in the matter limit ($w = 0$).

However, things are different when we consider transonic wiggly models \cite{CarterA,CarterB}. For Goto-Nambu strings, $T=\mu=\mu_0=const$, and $ \mu $ and $ T $ are the effective mass per unit length and tension of the string. Wiggly models obey $T\mu=\mu_0^2$, and have an equation of state
\begin{equation}
3w=\left(1+\frac{T}{\mu}\right)v^2-\frac{T}{\mu}\,.
\end{equation}
Notice that this does allow for the desired behavior: in the frozen limit we have
\begin{equation}
w=-\frac{T}{3\mu}
\end{equation}
while matterlike behavior ($w=0$) requires
\begin{equation}
v^2=\frac{T/\mu}{1+T/\mu}\,.
\end{equation}
The two regimes coincide in the tensionless limit $T/\mu\to0$: a frozen tensionless network has a matterlike equation of state. Moreover, for such a frozen network
\begin{equation}
\rho = \frac{\mu}{\xi^2} = \frac{\mu}{a^2} \propto a^{-3}\,,
\end{equation}
which implies that $\mu \propto a^{-1}$. 
This behavior is therefore quite different from that of an ordinary Goto-Nambu string network. In the context of 'standard' cosmological scenarios, one would need to invoke rather fine-tuned models to achieve it. On the other hand, it is certainly conceivable that such a behavior can ensue in cosmic (super)string models where some of the network's energy momentum gradually leaks into extra dimensions. This issue is beyond the scope of the present work, and its study is left for future work.

For our present purposes the relevant point is that, unlike the simplest Goto-Nambu strings, wiggly cosmic strings have a nonvanishing gravitational potential, $ \Phi $, which leads to the following gravitational force on a test particle \cite{Vilenkin}
\begin{equation}
\nabla\Phi=\frac{2G\left( \mu-T\right) }{r}\,,
\end{equation}
where $r$ is the distance from the center of the galaxy. Of course in the relevant tensionless limit $T\ll\mu$, so we will henceforth neglect the string tension. 

If a wiggly string threads the center of a spiral galaxy, the above contribution will add to the usual Newtonian one, so in the weak field limit the circular velocity is 
\begin{equation}\label{eq7}
v^{2}=\frac{GM}{r}+2G\mu.
\end{equation} 
This immediately shows that asymptotically the squared velocity is dominated by a constant term due to the string, so at least naively $ \sqrt{2G\mu} $ can be identified with the observational asymptotic velocities of spiral galaxies. 

We have analyzed several normal spiral galaxies with a range of total masses and different asymptotic velocities (in what follows we will show only four examples). We divided the galaxies in two groups: one group with the galaxies whose structure can be modeled assuming only a baryonic disk component (here represented by NGC3198 and NGC7331), and the other one with those where a bulge can not be ignored (here the galaxies NGC2841 and NGC2903).

\begin{figure}
\includegraphics[width=3in]{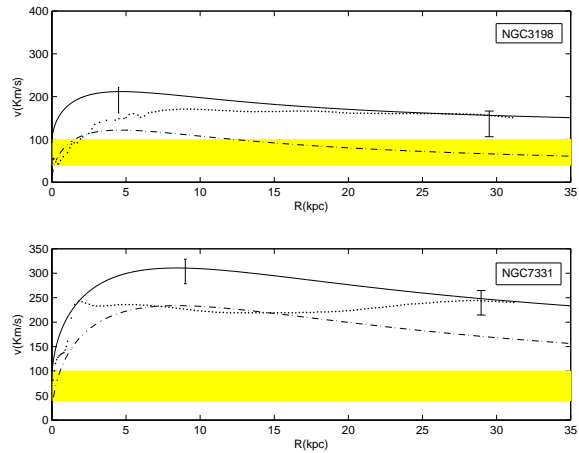}
\caption{\label{FigVibStab1} 
Rotation curves for two spiral galaxies with no bulge. Only a baryonic disc and a cosmic string threading each galactic center are assumed. The dotted (blue) line represents the observational data, and the dot-dashed one is the disk contribution. The solid (red) line is the total (disk $+$ string) rotation curve. The horizontal (yellow) bands and the bars represent, respectively, the ranges for $\sqrt{2G\mu}$ and for the total rotation curve discussed in the main text. For NGC3198 the best fit is obtained using $\sqrt{2G\mu} = 90\, {\rm km \, s^{-1}} $, while for NGC7331 that is achieved for $\sqrt{2G\mu} = 77\, {\rm km \, s^{-1}} $.}
\end{figure}

\begin{figure}
\includegraphics[width=3in]{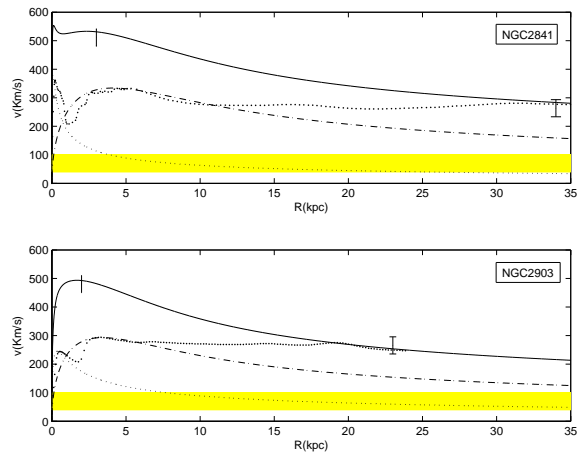}
\caption{\label{FigVibStab2} 
Same as for Fig. \protect\ref{FigVibStab1}, except that for these two galaxies one has to take into account a baryonic central bulge whose contribution to the total rotation curve (bulge $+$ disk $+$ string) is represented by the dashed line. Best fits were obtained with $\sqrt{2G\mu} = 90\, {\rm km \, s^{-1}} $ for NGC2841, and $\sqrt{2G\mu} = 40\, {\rm km \, s^{-1}} $ for NGC2903.}
\end{figure}

Considering a range for the parameter $ \sqrt{2G\mu} $ of between 40 and 100 ${\rm km \, s^{-1}}$, which in natural units corresponds to
\begin{equation}
10^{-9} \lsim G\mu \lsim 10^{-8}\,,
\end{equation}
one can easily show that it is possible to adjust the asymptotic velocities for most of normal spiral galaxies. For the galaxies shown in Figs. \ref{FigVibStab1} and \ref{FigVibStab2}, our best fits were obtained for significantly different values of $\sqrt{2G\mu}$, although they are all in the chosen range.  However, the constant term, due to the string, in Eq.\ (\ref{eq7}), also increases the velocities in the central region of the galaxies, which is in complete contradiction with observational data. As can be seen in both figures, the central region of the galaxies is never fitted. The discrepancy is less dramatic for NGC7331, but for this galaxy there is an intermediate region where no fit is possible.

\section{\label{LSS}Large scale structure}

Dark matter should be approximately homogeneous and isotropic on  cosmological scales, for otherwise that would yield strong (unobserved) signatures on the cosmic microwave background. If we take into account that the amplitude of the cosmic microwave background temperature perturbations generated by dark matter on scales of order $\sim H^{-1}_0/100$ cannot be much larger than $10^{-5}$ and assume that the power spectrum of density perturbations associated with the defects is white noise on scales larger than $\xi$, with the dispersion of order unity at the characteristic scale $\xi$ itself, we require $(100 H_0 \xi )^{3/2} \lsim 10^{-5}$ implying that $\xi \lsim 10 \, \rm kpc$.

Of course, an important aspect one should also consider is the compensation of defect perturbations on large scales. In standard defect models where the characteristic scale of the network is of the order of the Hubble radius, its main effect is to provide a cut-off to the spectrum of density fluctuations seeded by the defects at a scale of the order of the horizon (see for example \cite{Shoba,Robinson:1995xr,Cheung:1997tm}).

In this case, because compensation only acts on very large scales, the above calculation holds. However, if the defect network becomes frozen then compensation may become effective at a much smaller scale (still always larger than $\xi$). Nevertheless, at the characteristic scale, $\xi$, the defect energy density perturbation is of order unity and on scales larger than  $\xi$ the power spectrum of the defect energy density is white noise.  Consequently, even if we assume that compensation becomes effective on a scale $\xi_c$ with $\xi < \xi_c \ll H^{-1}$, the fact that during the matter era the growth factor is approximately equal to $z_{\rm eq} \sim 3.2\times 10^3$ would  lead to dramatic implications for the growth of small scale structures from very early on if $\xi \sim 1 \, \rm Mpc$. This would change the amplitude of small scale density fluctuations from an early stage strongly affecting the reionization history of the Universe \cite{Avelino:2003nn}.

In fact we expect that $\xi_c > H_{\rm eq}^{-1}$. On one hand, if the network freezes during the radiation era then in that epoch the compensation scale is comparable to the horizon if the decay products of the defect network move at relativistic speeds. On the other hand, even if this is not the case, perturbations in the radiation component propagate close to the speed of light and consequently one might expect that $\xi_c \sim H^{-1}$ during the radiation era. Deep in the matter era the fluctuations in the radiation component become less and less important and consequently if the source does not move then the comoving compensation scale should also not change much.

\section{\label{conc}Conclusions}

In this report we have considered the possible dynamical role of cosmic defects on scales significantly smaller than the horizon, In particular, we have discussed the possibility that a network of cosmic strings could provide the dark matter whose presence on galactic scales is suggested by the galactic rotation curves.

Our results show that even in principle only a nonstandard string network would have the key properties that could make the scenario viable. While no such scenarios are currently known (at least in quantitative detail) it is certainly conceivable that they may arise in cosmic (super)string models where some of the network's energy momentum gradually leaks into extra dimensions---such an energy leakage seems to be the only mechanism that could slow down the network to the required very low (effectively zero) velocities.

In any case, from the observational point of view the situation is quite clear. By adding a string contribution it is indeed quite easy to reproduce the asymptotic velocities for most of normal spiral galaxies, and the energy scale of the appropriate strings would in fact be the cosmologically interesting $G\mu\sim10^{-8}$. However, any such string contribution will also increase the velocities in the central region of the galaxies, which is in complete contradiction with observational data. We thus conclude that although defects can still play a number of interesting astrophysical roles on galactic scales, they are unable to shed light on the dark matter problem.

%%%%%%%%%%%%%%%%%%%%%%%%%%%%%%%%%%%%%%%%%%%%%%%%%%%%%%%%%%%

\begin{acknowledgments}
We thank Catarina Lobo and Y. Sofue for valuable advice and enlightening comments. The work of C.M. is funded by a Ciencia2007 Research Contract.
\end{acknowledgments}
\bibliography{stringsCDM}
\end{document}